\documentclass[showpacs,preprintnumbers,english,amsmath,amssymb]{revtex4-1}
\usepackage[T1]{fontenc}
\usepackage[latin1]{inputenc}
\usepackage{graphicx}
\usepackage{amssymb}
\usepackage{verbatim}
\usepackage{epic,eepic}
\usepackage{babel}

\begin{document}

\title{Virial Expansion of the Nuclear Equation of State}

\author{Ruslan Magana$^{a,b)}$, Hua Zheng$^{a)}$ and Aldo Bonasera$^{a,c)}$}
\affiliation{
a)Cyclotron Institute, Texas A\&M University, College Station, TX 77843, USA\\
b)Instituto de Ciencias Nucleares, Universidad Nacional Aut\'onoma de M\'exico, 
A.P. 70-543, 04510 M\'exico, D.F., M\'exico\\
c)Laboratori Nazionali del Sud, INFN, via Santa Sofia, 62, 95123 Catania, Italy.}




\begin{abstract}
We study the equation of state (EOS) of nuclear matter as function of density. We expand the  energy per particle (E/A) of symmetric infinite nuclear matter in  powers of the
 density to take into account 2,3,$\ldots$,N-body forces. New EOS are proposed by fitting ground state properties of nuclear matter (binding energy, compressibility and pressure) and assuming that at high densities a second order phase transition to the
Quark Gluon Plasma (QGP) occurs.  The latter phase transition is due to symmetry breaking at high density from nuclear matter (locally color white) to the QGP (globally color white).  In the simplest 
implementation of a second order phase transition we calculate the critical exponent $\delta$ by using Landau's theory of phase transition.  We find $\delta=3$.
  Refining the properties of the EOS near the critical point gives $\delta=5$ in agreement with experimental results. We also discuss some scenarios for the EOS at finite temperatures.
\end{abstract}


\maketitle

\section{Introduction}
In recent years, the availability of new heavy-ion accelerators which are 
capable of accelerating ions from a few MeV/nucleon to GeV/nucleon has fueled a new field of research loosely referred to as Nuclear Fragmentation. The characteristics of the fragments produced 
depend on the beam energy and the target-projectile combinations which can be externally controlled. Depending on the beam energy, hard photons, pions, kaons and so on can be  produced as well \cite{1,2}.
Fragmentation  experiments could provide informations about the nuclear matter properties and constrain the EOS of nuclear matter.  A 'conventional' EOS provides only limited information about the nuclear matter: 
the static thermal equilibrium properties\cite{Csernai}. In heavy ion collisions non-equilibrium processes are very important, thus nuclear transport properties will play an equally important role.  If we want to study
 the EOS at high densities and high temperatures, we have to rely on theoretical estimates as well. The low density behavior of nuclear matter determines the observables mechanism of the final expansion stage in a collision before the break up. 
 The low density part of the nuclear EOS  is directly related to the final fragmentation, nuclear compressibility, momentum dependence, etc \cite{Csernai}.
After an energetic nucleus-nucleus collision in the 100 MeV-4 GeV/nucleon beam energy region, many light nuclear fragments, a few heavy fragments and a few mesons (mainly pions) are observed. 
Thus the initial kinetic energy of the projectile leads to the destruction of the ground state of nuclear matter and converts it into dilute gas  of fragments which then loses thermal contact during the break-up or freeze-out stage.

One of the standard methods to explore the nuclear EOS is within the  framework of the mean field theory. It starts out with a Langrangian including the nucleon field $\Psi$, a scalar meson field $\phi$, and a vector meson field $V_\mu$. Customarily the contribution of the scalar field is described by a quartic polynomial.
 
 From conventional nuclear physics we know that there is a stable equilibrium state at the normal nuclear density $\rho_0=0.145-0.17 fm^{-3}$\cite{3,4}
  with a compressibility  in the range of $K=180-240$ MeV \cite{5} and a binding energy of  15-16 MeV/nucleon\cite{shlomo}. In this work we take $\rho_0=0.165 fm^{-3}$ and $K=225$ MeV based on
  the condition that the mean field potential has a minimum at normal nuclear density.  With increasing density, the effects of  N-body correlations become more and more important.  This is especially true near a phase transition.  
  Furthermore, nucleons are not elementary particles but they are made of quarks and gluons, thus N-body forces are expected to be stronger at high densities where
the nucleon wave functions strongly overlap.  We can take these features into account by 
expanding the EOS in powers of the density as customary in the virial expansion of any EOS\cite{8}.

 Finally we discuss some properties at finite temperatures assuming either a classical gas or a quantum Fermi systems.  We show
that at the densities and temperatures of interest the classical approximation is not valid.  This is at variance with many experimental and theoretical results in heavy ion collisions near the Fermi energy \cite{albergo,pocho,15} which assume the classical approximation to be valid. 
Quantum corrections have been recently extensively discussed in \cite{15a} for cluster formation in a low density expanding nuclear system.

\section{Nuclear Equation Of State}

For a system interacting through two body forces having a short-range repulsion and a longer-range attraction the EOS resembles a Van Der Waals one. This is indeed the case for nuclear matter \cite{1,2,Csernai,10}.

A popular approach is to postulate an equation of state which satisfies known properties of nuclei \cite{shlomo}. The equation for energy per particle is:

\begin{equation}
 E/A=22.5\tilde \rho^{\frac{2}{3}}+\frac{A}{2}\tilde \rho+\frac{B}{\sigma+1}\tilde\rho^\sigma
\label{2.1}
\end{equation}
where  $\tilde \rho=\rho/\rho_0$ and $\rho_0$  is the normal nuclear density.  The first term of  eq. (\ref{2.1})  refers to the kinetic energy of a  Fermi gas.  The other terms are due to potential interactions and correlations. To generalize at finite temperature we could use a classical approximation giving an EOS in the form \cite{10}.
\begin{equation}
 P=\rho^2\frac{\partial (E/A)}{\partial \rho}+\rho T
\label{2.2}
\end{equation}
In the following we will test the validity of such an approximation.

The requirement of causality provides several theoretical constraints on the EOS \cite{11} at high densities and limits the choice of the functional form of the compressional energy that can be used in phenomenological EOS(see Fig. \ref{Fig4}). 
Very stiff equations of state may lead to superluminal speed of sound (see ref. \cite{12}). However, this is not a problem if the acausality occurs in a region of the phase diagram where matter is in the mixed or plasma phase, because  the phase transition softens the EOS. 
There are some phenomenological parameterizations of the specific energies like Linear Quadratic, Sierk-Nix or Grant-Kapusta  which are acausal at sufficiently high densities\cite{Csernai}. 
Fortunately the acausality occurs well within or beyond the mixed or plasma phase for all the parameterization except the Quadratic. The basic parameter is the (isothermal) compressibility, which is defined as
\begin{equation}
 K=9\frac{\partial P}{\partial\rho}\bigg|_{\rho=\rho_0,T=0}
\label{2.3}
\end{equation}

The three parameters $A$, $B$ and $\sigma$ in  eq. (\ref{2.1}) are determinated by using the conditions of pressure equal zero,  the binding energy of $E/A=-15$MeV and finally the compressibility is of order of 200MeV (as inferred from the vibrational frequency of the giant monopole resonance\cite{5}), at $\rho=\rho_0$.
 Using these conditions, we get $A=-356$MeV, $B=303$MeV and $\sigma=7/6$, we will refer to this EOS as CK200 (Conventional, K=200 MeV).

Now if we modify this approach in accordance to \cite{13} the compressibility condition is substituted by the mean field potential having a minimum at the ground state density:
\begin{equation}
 U(\rho)=A(\tilde \rho)+BA(\tilde\rho)^\sigma
\label{2.4}
\end{equation}
where
\begin{equation}
 \frac{\partial U(\rho)}{\partial \rho}\bigg|_{\rho=\rho_0}=0
\label{2.5}
\end{equation}
which means that the compressibility at $\rho=\rho_0$ is equal to $K=225$ MeV then we have the conditions
\begin{equation}
\begin{array}{lll}
a)&E/A=-15\text{MeV}\\
b)&P=0&\text{at}\hspace{0.5cm}\rho=\rho_0\\
c)&K=225\text{MeV}
\end{array}
\label{2.6}
\end{equation}
Solving these equations, we get $A=-210$ MeV, $B=157.5$ MeV and $\sigma=4/3$, we will refer to this EOS as  CK225.

The form of the EOS is a very delicate subject. For a nuclear system we expect to see a liquid-gas (LG) 
phase transition at a temperature of the order of 10 MeV and at low density. Under these conditions we can assume that nuclear matter behaves like a classical ideal gas eq. (\ref{2.2}) however this is just our ansatz. 
 Actually in this work we will compare the theoretical behavior of the EOS assuming classical ideal gas and Fermi gas.

\begin{figure}
\centering
\includegraphics[width=0.9\columnwidth]{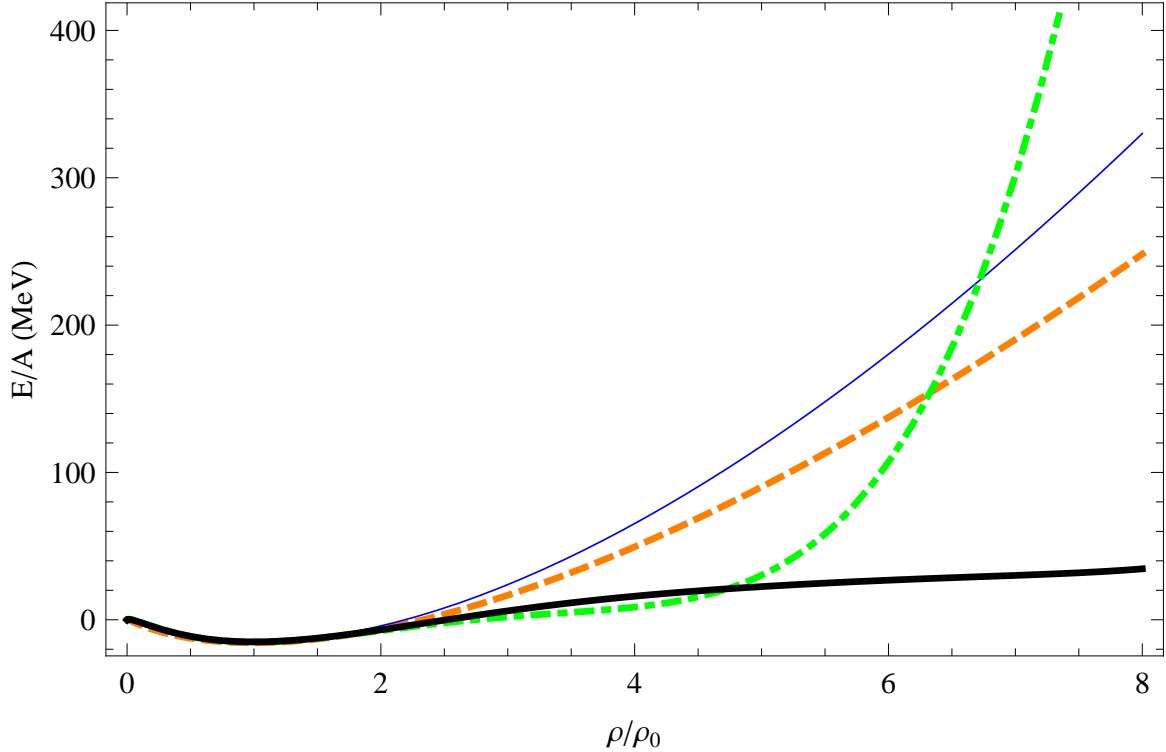} 
\caption[]{The energy per particle of nuclear matter as a function of density using different density-dependent interactions in comparison with the 'conventional' formulation. 
The parameter values are given in Table~\ref{table1}.  Symbols: CK225-thin solid line, CK200-dashed line, CCS$\delta$3-dashed dotted line and  CCS$\delta$5-thick line.}
\label{Fig1}
\end{figure}

In order to calculate the critical point, we will impose the conditions that the first and second derivative of eq. (\ref{2.2}) respect to density are equal to zero therefore we can obtain the critical temperature and density, i.e. $T_c=9$ MeV and $\rho_c=0.3\rho_0$. Such values are in some agreement with experimental results \cite{10}. 
However, we notice that in order for the classical approximation to be valid, the ratio of the temperature to the Fermi Energy, $\epsilon_f(\rho)=36(\rho/\rho_0)^{2/3}MeV$, should be much larger than one. For the values above we get $T_c/\epsilon_f=9/16<1$ which shows that we are still in a quantum regime.

The validity of eq.(1) is restricted to densities close to the ground state value.  In fact no further constraints are imposed so far  for larger densities.  Such constraints should come from experimental data in heavy ion collisions and properties of heavy stars.  Those data, if available, do not give directly
a constraint on the EOS but must be filtered through model calculations.  The models in turn need some form of EOS. We propose a new equation for the energy per particle which could be used in microscopic calculations:
\begin{equation}
 E/A=\tilde \varepsilon_f\tilde\rho^\frac{2}{3}+\sum_{n=1}^k\frac{A_n}{n+1}\tilde \rho^n
\label{2.7}
\end{equation}
where the first term refers to the average kinetic energy of a free Fermi gas with $\tilde \varepsilon_f=3/5 \varepsilon_f= 22.5 MeV$, the other terms are due to  potential interactions and correlations. The $n=1$ term  is obtained by taking into account  the interaction between pair of particles, and the subsequent terms must involve the interactions between groups of three, four, etc., particles. 
The coefficients $A_n$ in the expansion, eq. (\ref{2.7}) are called first, second,  third, etc., virial coefficients\cite{8}. 

Let's start considering three body forces $O(\rho^3)$ and we assume that at the normal density and zero temperature the energy of the ground state is -15 MeV and the pressure is zero. In addition we assume that the mean field potential has a minimum at normal density or, equivalently $K=225$ MeV.
If we do this, we have three conditions and three equations so we can solve the corresponding set of equations. Unfortunately, the solution has no physical meaning because the energy diverges to minus infinity when the density  approaches infinity, see table I.  

For the fourth order of our expansion  the eq. (\ref{2.7}) takes the form
\begin{equation}
 E/A=\tilde\varepsilon_f\tilde\rho^\frac{2}{3}+\frac{A_1}{2}\tilde\rho+\frac{A_2}{3}\tilde \rho^2+\frac{A_3}{4}\tilde\rho^3+\frac{A_4}{5}\tilde\rho^4
\label{2.8}
\end{equation}

Here we assume the symmetry breaking at high density, from nuclear matter (locally color white) to the  QGP (globally color white), this possibly gives a second-order phase transition\cite{16,17}.

In accordance to the conditions given in eq. (\ref{2.6}) we can add two extra constraints based on the conditions  of matter close to the critical density of a second-order phase transition at  T=0.

\begin{table}[t]

\centering
\caption{Values of the parameters of the EOS's of nuclear matter.}
\begin{tabular}{cccllllllllll}
\hline\noalign{\smallskip}
\textbf{Interactions}&$A_1$&$A_2$&$A_3$&$A_4$&$A_5$&$A_6$&$\sigma$\\
&\scriptsize[MeV]&\scriptsize[MeV]&\scriptsize[MeV]&\scriptsize[MeV]&\scriptsize[MeV]&\scriptsize[MeV]\\
\hline\noalign{\smallskip} 
\textbf{CK200}&-356&303&&&&&7/6 \\\noalign{\smallskip}
\textbf{CK225}&-210&157.5&&&&&4/3 \\\noalign{\smallskip}
\textbf{3}&-135&112.5&-30 \\\noalign{\smallskip}
\textbf{CCS$\bf\delta$3}&-136.89&120.99&-41.32&4.72 \\\noalign{\smallskip}
\textbf{5}&-137.59&124.41&-46.51&7.67&-0.47 \\\noalign{\smallskip}
\textbf{CCS$\bf\delta$5}&-137.96&126.25&-49.46&9.55&-0.92&0.0035 \\
\hline\noalign{\smallskip}
\multicolumn{8}{l}{\scriptsize  The 'Conventional' EOS corresponds to eq. (\ref{2.1}) where the the ground}\\	
\multicolumn{8}{l}{\scriptsize  state binding energy  E$_{\text{min}}$=-15 MeV and the compressibility at normal}\\
\multicolumn{8}{l}{\scriptsize  nuclear density K=200-225 MeV and the remaining EOS's correspond to }\\
\multicolumn{8}{l}{\scriptsize  eq. (\ref{2.7}) with E$_{\text{min}}$=-15 MeV and K=225 MeV respectively.}\\
\end{tabular}
\label{table1}
\end{table}
At the critical point the first and second derivative of the pressure with respect to  density are equal to zero. Then we have five constrains:

\begin{equation}
\begin{array}{lll}
a)&E/A=-15\text{MeV}\\
b)&P=0&\text{at}\hspace{0.5cm}\rho=\rho_0\\
c)&K=225\text{MeV}\\
d)&\frac{\partial P}{\partial\rho}\bigg|_{\rho=\rho_c}=0\\
e)&\frac{\partial^2 P}{\partial\rho^2}\bigg|_{\rho=\rho_c}=0
\end{array}
\label{3.1}
\end{equation}

Solving these equations we get the values of $A_1..A_4$ (see table \ref{table1}).  Therefore we obtain the critical point $\tilde \rho_c=2.9354$, we refer to this EOS as CCS$\delta$3. This could be the critical point for a second-order phase transition to the QGP at T=0.

In order to include higher order terms  we assume that $\frac{\partial^n P}{\partial\rho^n}\bigg|_{\rho=\rho_c}=0$ where $n=1,\ldots,4$  (this assumption will become clear later on) then we solve the resulting nonlinear system of many unknown variables to get all $A_n$ and therefore the critical point $\tilde \rho_c$ at T=0. 
Here we are taking into account the interaction between pairs of particles and interactions between groups of three, four, five and six particles.

For interactions up to five particles $O(\rho^5)$  where $\frac{\partial^n P}{\partial\rho^n}\bigg|_{\rho=\rho_c}=0$ with n=1-3, the nuclear EOS diverges negatively again as in the  $O(\rho^3)$ case.

If we take more variables in our expansion, up to  sixth order we get:
\begin{equation}
 E/A=\tilde\varepsilon_f\tilde\rho^\frac{2}{3}+\frac{A_1}{2}\tilde\rho+\frac{A_2}{3}\tilde \rho^2+\frac{A_3}{4}\tilde\rho^3+\frac{A_4}{5}\tilde\rho^4+\frac{A_5}{6}\tilde\rho^5+\frac{A_6}{6}\tilde\rho^6
\label{3.2}
\end{equation}

Imposing the conditions that the fourth order derivative of the pressure vanishes as well, gives the values of the parameters reported in Table I and a critical density $\tilde \rho_c=5.2$, which we refer as CCS$\delta$5. 

Now we can try to apply our equations of state in the high-density domain. A quite promising result is shown in Fig. \ref{Fig1}, where we compare the different EOS at  T=0.
\begin{figure}
\centering
\includegraphics[width=0.9\columnwidth]{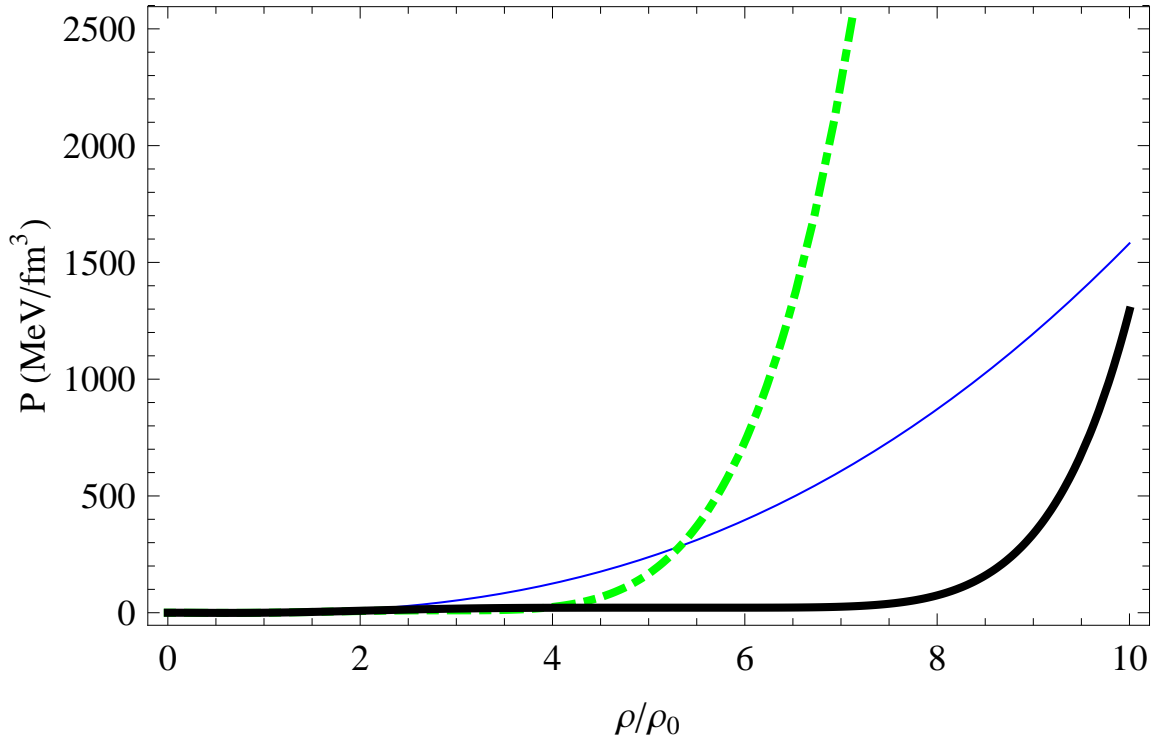} 
\caption[]{The pressure per particle of nuclear matter as a function of density using different density-dependent interactions. Symbols:  CK225-thin line, CCS$\delta$3-dashed dotted line and  CCS$\delta$5-thick line.
}
\label{Fig2}
\end{figure}
The different EOS are very similar near and below the ground state density of nuclear matter while they differ greatly, as expected at higher densities.  In particular the EOS softens because of the
assumed QGP phase transition.

The pressure is shown in Fig. \ref{Fig2} at temperature T=0 with a comparison with the conventional EOS.  At $\rho=\rho_0$ the pressure is zero and increases largely for the 'conventional' EOS.

\begin{figure}
\centering
\includegraphics[width=0.9\columnwidth]{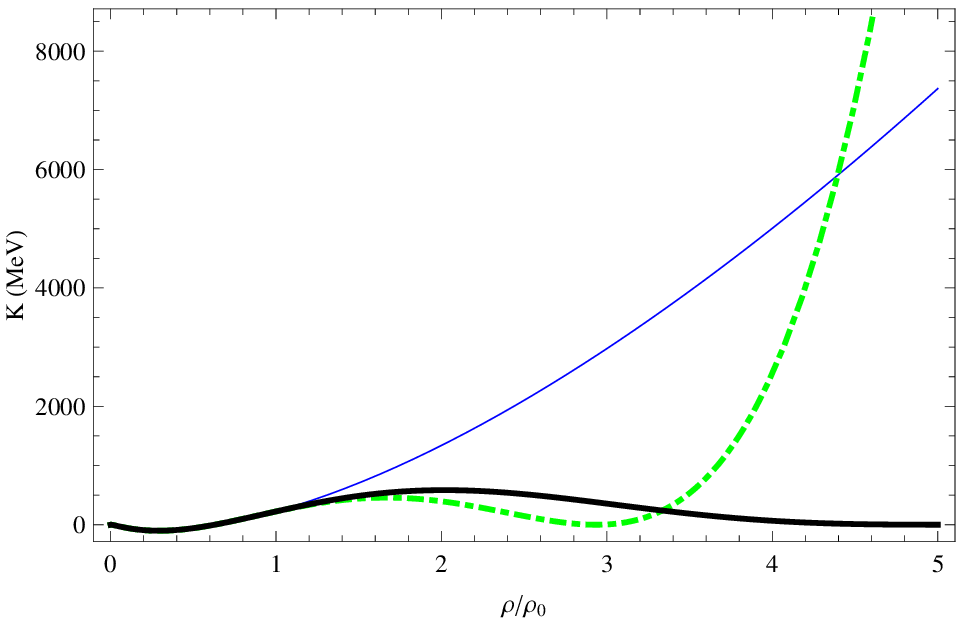} 
\caption[]{The compressibility per particle of nuclear matter as a function of density using different density-dependent interactions. Symbols as in Fig.2.}
\label{Fig3}
\end{figure}
We proceed now to the study of the compressibility. We are assuming that at $\rho=\rho_0$,  K=225 MeV (apart CK200) as is shown in Fig. \ref{Fig3}. A negative compressibility (which indicates an instability region) is obtained at subnuclear densities only, while it becomes
  zero at the critical point for the QGP phase transition.

A negative compressibility gives an imaginary speed of sound since for a particle of mass m \cite{Landau-Vol6}:

\begin{equation}
v_c=\sqrt{\frac{1}{m}\frac{\partial P}{\partial \rho}}.
\end{equation}

Such a quantity is plotted in Fig.4 as function of density for the different EOS.  Two phenomena are worth noticing.  First the speed of sound becomes larger than the speed of light for all the EOS (excluding CCS$\delta$5) for $\rho > 4.5\rho_0$.  The CCS$\delta$5 EOS gives a superluminal speed of sound
at almost twice such a density, well in the region of the QGP. 

The second property is that the speed of sound becomes imaginary in the instability region, thus a discontinuity is shown in Fig. 4 below normal nuclear matter density.  The small discontinuity observed near the critical density of CCS$\delta$5 is due 
to the numerical solution of the set of equations used to determine the values of the coefficients reported in Table I.  We can have a better view of the critical region by zooming on the density as in Fig. 5.  Now the discontinuities at lower densities due to the LG 
 phase transition are visible for three EOS. The instability region
is very much the same for all the EOS having the same compressibility, which shows that such a region is mainly determined by the ground state properties of the EOS and not by the assumed functional form.  As we will show later also the critical point for the LG is the same when the compressibility is the same.  At higher densities,
the speed of sound becomes zero (discontinuous) at the critical densities of the fourth and sixth order EOS respectively.
\begin{figure}
\centering
\includegraphics[width=0.9\columnwidth]{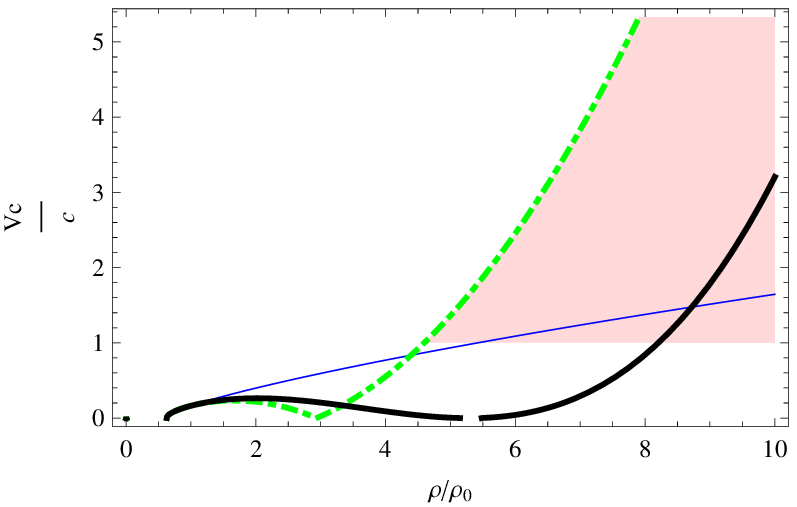} 
\caption[]{The speed of sound of nuclear matter as a function of density. In the shaded region the  principle of causality is broken. Symbols as in Fig.2.}
\label{Fig4}
\end{figure}

\begin{figure}
\centering
\includegraphics[width=0.9\columnwidth]{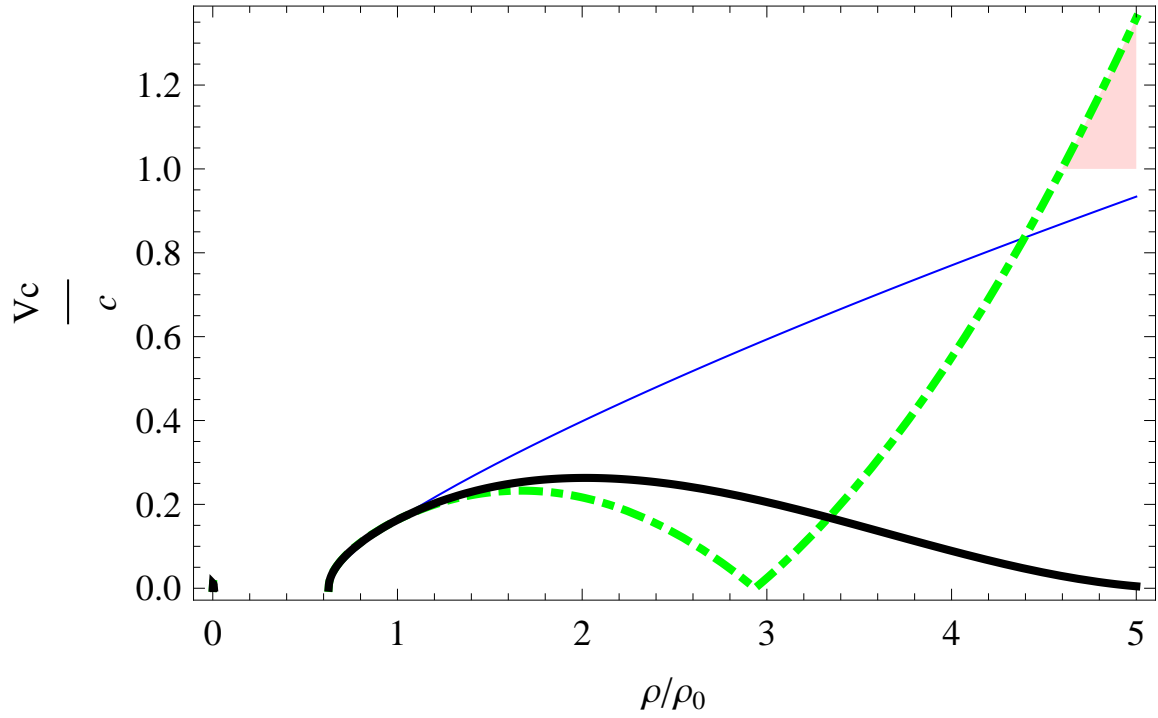} 
\caption[]{The speed of sound  of nuclear matter as a function of density. Symbols as in Fig.2.}
\label{Fig5}
\end{figure}

\begin{figure}
\centering
\includegraphics[width=01.\columnwidth]{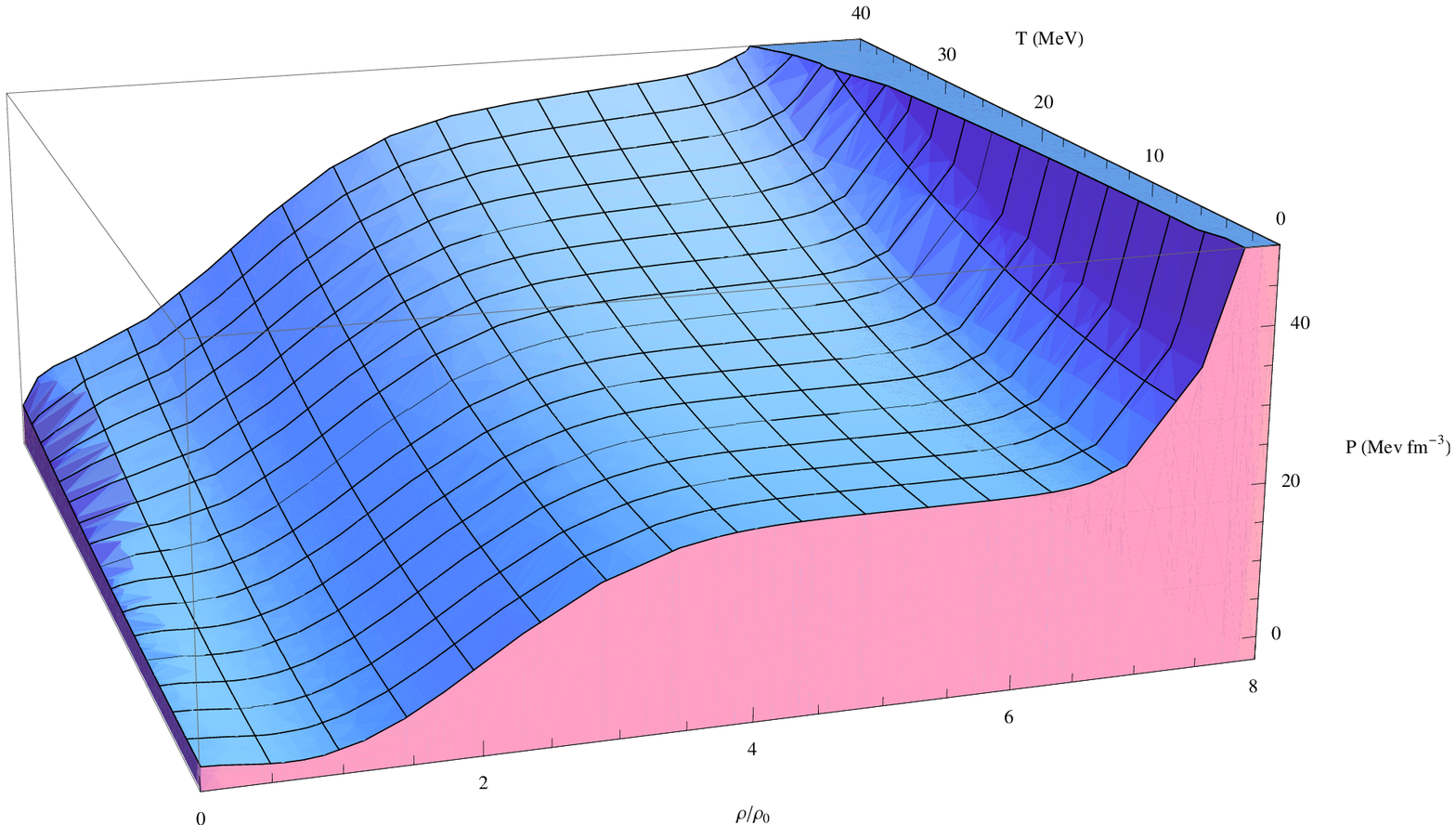} 
\caption[]{Equation of state surface for a nuclear system with a second-order phase transition to the QGP, CCS$\delta$5. The pressure is reduced by the phase transition  near the critical region. The pressure increases again at very high densities in the QGP phase.}
\label{Fig6}
\end{figure}
\section{Finite Temperatures}
In order to study the properties of the EOS at finite temperatures we need to go beyond the classical approximation.  A simple functional form could be obtained using a Fermi gas expansion\cite{8,10,15}.


The region of validity of the Fermi gas model is related to the ratio of the temperature to the Fermi energy $T/\varepsilon_f(\rho)$. In this scenario eq. (\ref{2.7})  takes the form:
        
\begin{equation}
E/A=\tilde \varepsilon_f \tilde \rho^{\frac{2}{3}}+\sum_{n=1}^k\frac{A_n}{n+1} \tilde \rho^n+a_0\tilde\rho^{-\frac{2}{3}}T^2
\label{3.6}
\end{equation}
where the  level density parameter $a_0=1/13.3$ MeV$^{-1}$. Correspondingly the pressure takes the form\cite{8}:
\begin{equation}
 P=\rho_0\left[15\tilde \rho^{\frac{5}{3}}+\sum_{n=1}^k\frac{nA_n\tilde \rho^{(n+1)}}{(n+1)}+\frac{2}{3}a_0\tilde \rho^{\frac{1}{3}}T^2\right]
\end{equation}

\begin{figure}
\centering
\begin{tabular}{ccc}
\includegraphics[scale=0.63]{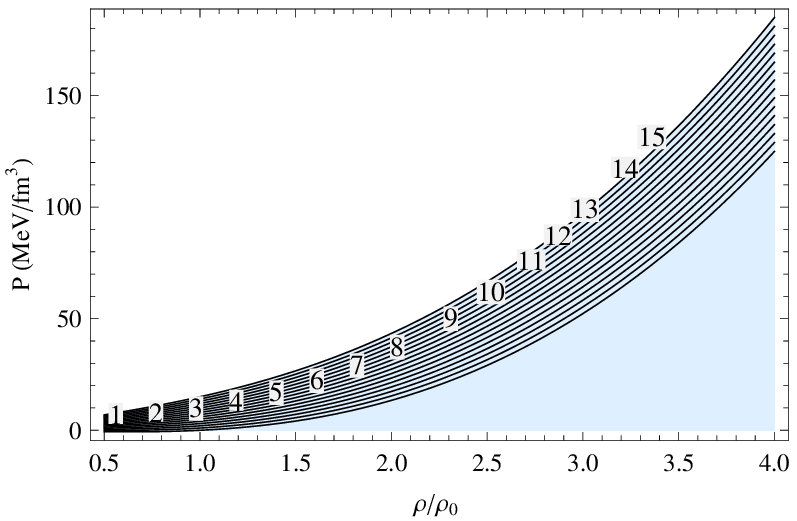}\\\includegraphics[scale=0.6]{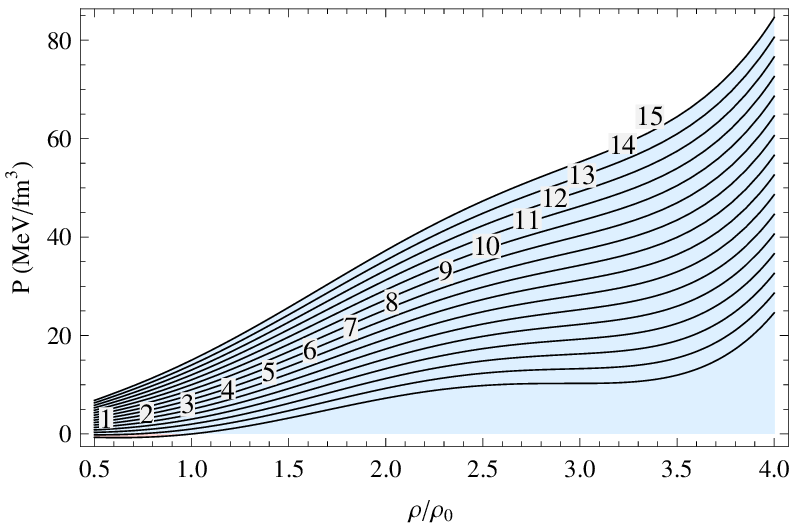}\\\includegraphics[scale=0.6]{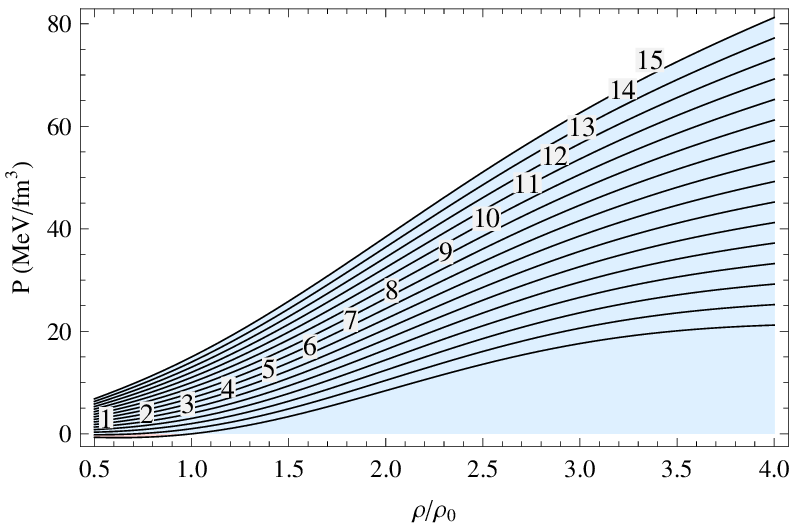}
\end{tabular}
\caption{
The pressure per particle of nuclear matter as a function of density at some temperatures. Here we are assuming  that nuclear matter behaves like a classical ideal gas. The EOS are respectively from top to bottom: CK225, CCS$\delta$3 and CCS$\delta$5.
}
\label{Fig7}
\end{figure}

\begin{figure}
        \centering
        \begin{tabular}{ccc}
        \includegraphics[scale=0.7]{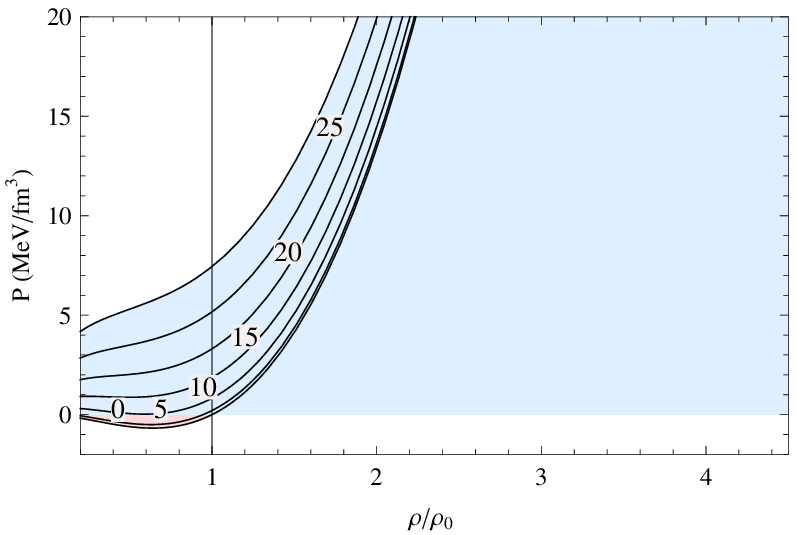}\\\includegraphics[scale=0.6]{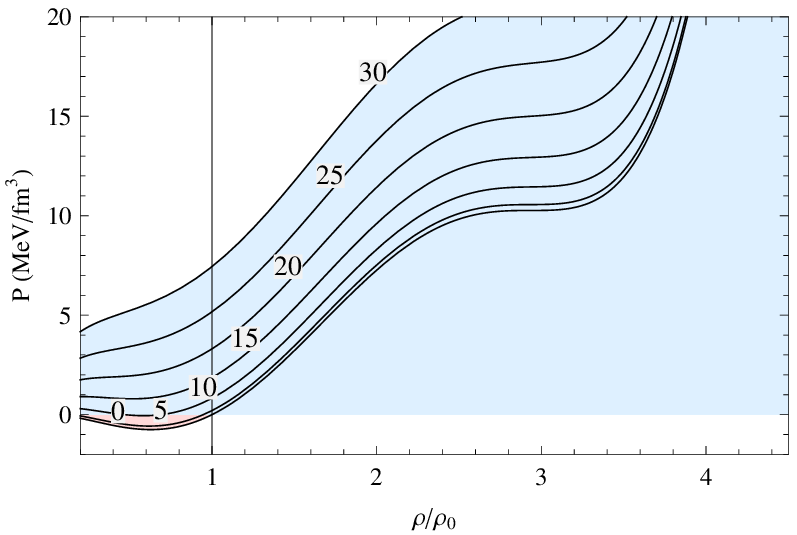}\\\includegraphics[scale=0.6]{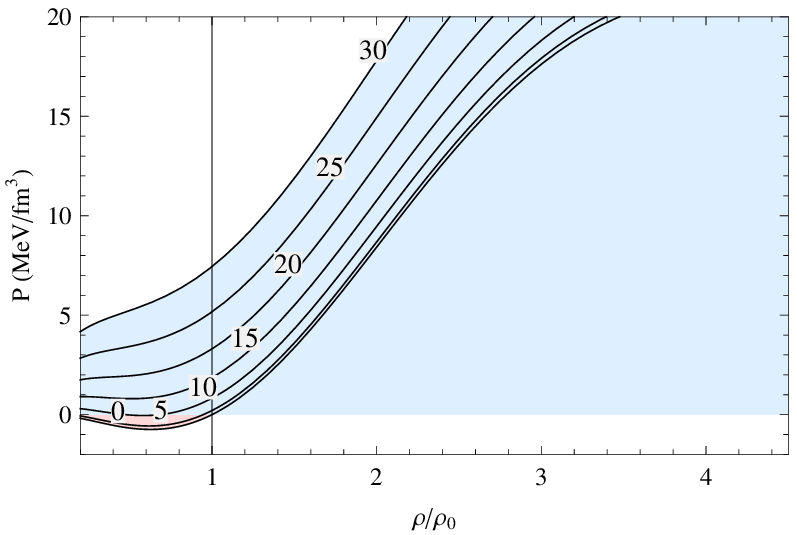}
        \end{tabular}
\caption{Same as Fig.7 but for  a Fermi gas.}
\label{Fig8}
\end{figure}

\begin{figure}        
        \centering
        \begin{tabular}{ccc}
        \includegraphics[scale=0.6]{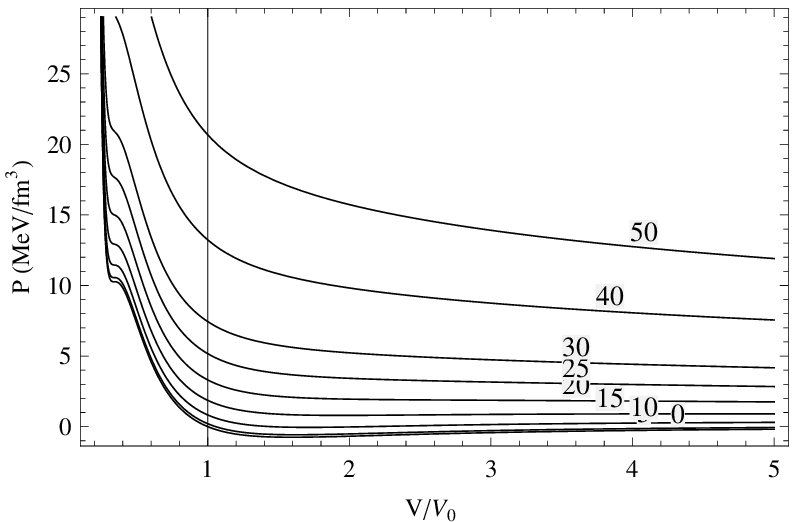}\\\includegraphics[scale=0.6]{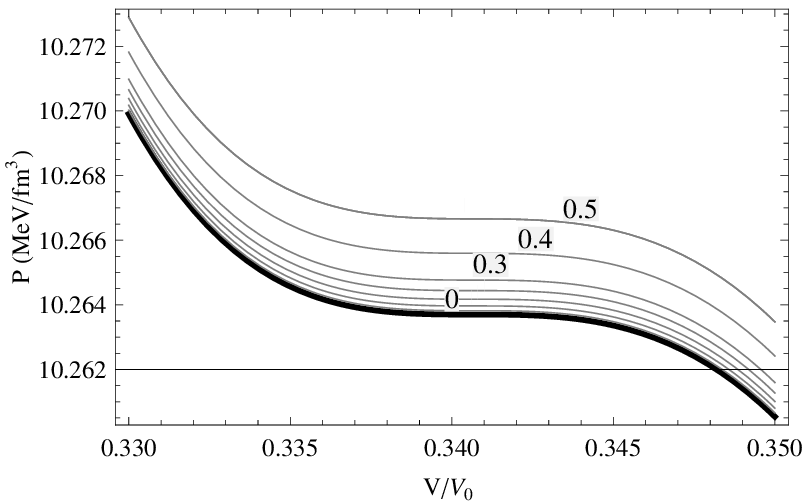}\\\includegraphics[scale=0.6]{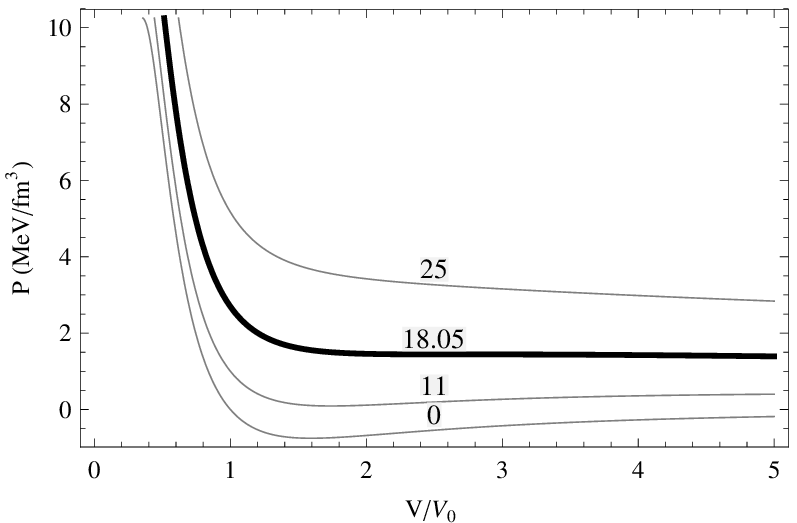}
        \end{tabular}
        \label{Fig9}
\caption{Behavior near to the critical point for $O(\rho^4)$; the pressure per particle of nuclear matter as a function of volume at some temperatures. Here we are assuming that nuclear matter behaves like a Fermi gas.
}
    \end{figure}

\begin{figure}        
        \centering
        \begin{tabular}{ccc}
        \includegraphics[scale=0.6]{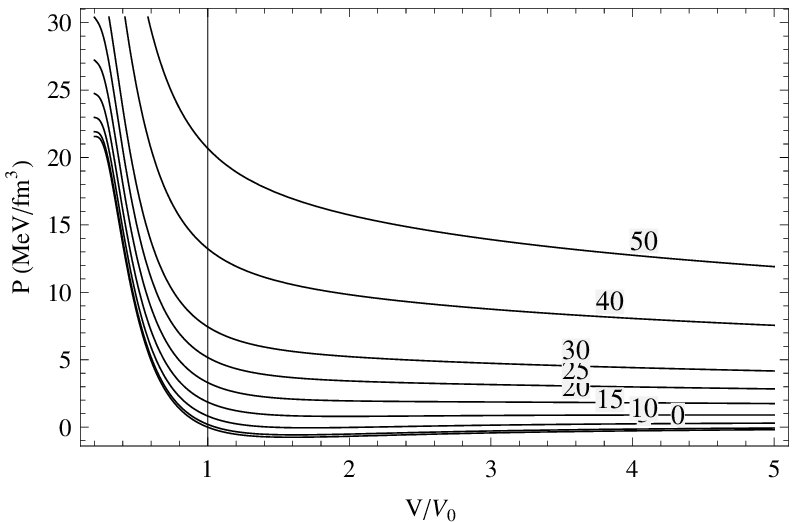}\\\includegraphics[scale=0.6]{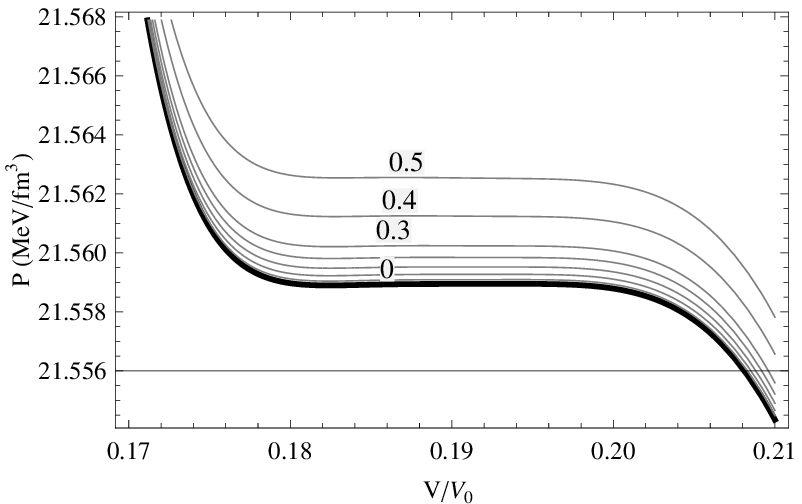}\\\includegraphics[scale=0.6]{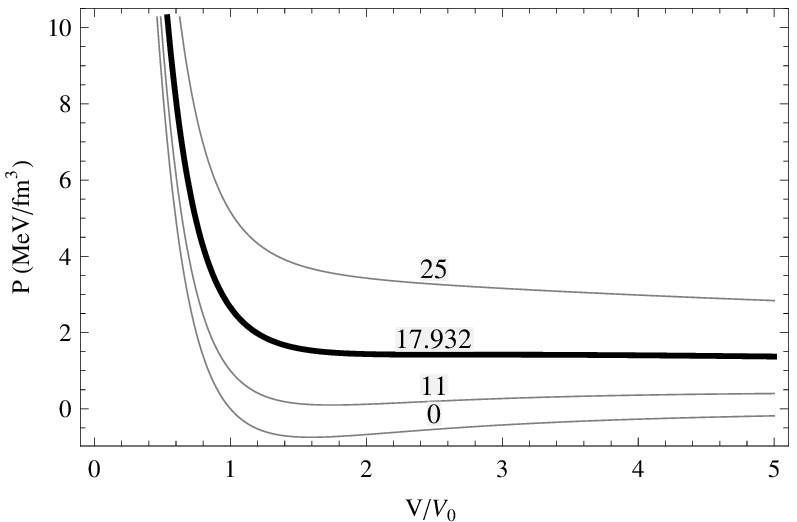}
        \end{tabular}
        \label{Fig10}
\caption{Behavior near to the critical point  for $O(\rho^6)$; the pressure per particle of nuclear matter as a function of volume at some temperatures. Here we are assuming that nuclear matter behaves like a Fermi gas.
}
    \end{figure}

The properties near the critical temperature and density can be obtained for the various EOS imposing the constraints similar to those given in eq.(9d),e)).
For the case of fourth order $O(\rho^4)$ the critical temperature and critical density for the LG phase transition has the values $T_c=18.05$ MeV  and $\rho_c=0.3724\rho_0$ respectively.
In this approach we are assuming interactions between pairs of particles and between groups of three and four particles which could be fully justified by the fact that nucleons are made of quarks and gluons.
 For higher order terms, $O(\rho^6)$ the critical temperature and density are $T_c=17.932$ MeV and $\rho_c=0.3717\rho_0$. Similar values are obtained for the CK225 EOS. We notice that the critical values of the LG do not change much by using different EOS.
 In particular the critical density seems to be almost independent of the assumed classical or quantum statistical properties of nuclear matter.  On the other hand, the critical T changes almost of a factor two when going from the classical to the quantum approximation.  We have already noticed that
 the ratio of T to the corresponding Fermi energy at the critical density was smaller than one when using the classical approximation.  If we perform the same calculation for the quantum case we get $T_c/\epsilon_f=18/18.6\approx1$, which implies that higher order terms must be considered in the expansion of
 the Fermi gas pressure (or energy per particle) but we are $\it{not}$ in the classical regime.  A detailed calculation at finite T properly taking into account the Fermi statistics is beyond the scope of this work but we would expect a slight reduction of the temperature as compared to the value given above.
 We notice again that values obtained experimentally for the critical T and $\rho$ suggest that we are in the quantum regime while the methods used to obtain them are purely classical \cite{albergo,pocho,15}, but \cite{15a}.  This is analogous to what we have done above, eq.(2), using a classical approximation for finite T.  We expect that
  properly taking into account quantum statistics will dramatically change the values obtained experimentally near the critical point.
  
  In the quantum  approximation considered above we can obtain the $P(\rho, T)$, the result is given in Fig.6 for the CCS$\delta$5 case.  This region is very similar for all the EOS for a given compressibility (K=225 MeV in our case).
  At higher T and/or $\rho$ the pressure flattens because of the QGP phase transition and increases again at higher densities.  
    
We can compare  the differences between different EOS more in detail in the purely classical, Fig.\ref{Fig7} and quantum approximation, Fig.\ref{Fig8}. We notice the dramatic differences between the two cases especially at the lowest densities and T where quantum effects are stronger.  For completeness we show in Figs.9 and 10 the same cases but as function of reduced volume.

\section{ Critical Phenomena}
The term critical phenomena refers to the thermodynamic behavior  of a system near the critical temperature  of a second order phase transition. A simple understanding of this phenomenon can be obtained in the framework of Landau's theory\cite{8}.  
In such a framework it is assumed that near the critical point of a second order phase transition, the relevant degree(s) of freedom reduce to a few (order parameters) which reflect  basic invariance properties of the Hamiltonian.
  Below the critical point such  invariance(s) is spontaneously broken and it is restored above the critical point, this means that
the order parameter is non zero only below the critical point.  Some examples are the magnetization in a ferromagnetic system, or the difference in densities between a gas and its liquid in a liquid-gas phase transition.  In Landau's approach one assumes that there exist a free energy $\Psi$ near the critical point that depends on the order parameter and its conjugate field.
In general $\Psi=E/A-TS$, where S is the entropy.  Near the critical point we can expand the free energy in terms of the density at the critical temperature (which is zero for our case). Defining the order parameter $\eta=V-V_c$ i.e. the distance from the critical volume, we get
\begin{equation}
\Psi=E(V)/A\approx\sum_n\frac{1}{n!}\frac{\partial^{(n)}E(V)}{\partial V^{(n)}}\bigg|_{V_c}(V-V_c)^{(n)}
\end{equation}
n=0,1...N. We define  an external or conjugate field $P(V_c)=-\partial E/\partial V|_{V=V_c}$. If we stop our expansion at fourth order $n=4$, and notice that  in the neighborhood of the critical point the conditions of the second order phase transition are $P'(V_c)=0$ and $P''(V_c)=0$ at $T=0$ . Thus only n=0 (a constant),1 and 4 survive in the expansion above.
Imposing the free energy to have a minimum in presence of a conjugate field gives\cite{8}:

\begin{equation}
\eta\sim P^{\frac{1}{\delta}}\sim P^{\frac{1}{3}}
\end{equation}
Therefore $O(\rho^4)$  has the critical exponent $\delta=3$ which corresponds to the 'mean field' or 'classical' value\cite{8}. Experimentally the value of  $\delta=4-5$\cite{8} is found. It is now evident how to go beyond the 'mean field' value using the arguments given above.  In particular if we impose $P'''(V_c)=0$ and continue the expansion in eq.(14) to
$n=5$ we easily get $\delta=4$.  However the resulting fifth order coefficient in the virial expansion of the EOS is negative, see table I, thus this case is unphysical. Finally expanding to $n=6$ and imposing $P''''(V_c)=0$ gives $\delta=5$ and the fitting parameters reported in table I for the CCS$\delta$5 EOS.
Notice again that our formulation is perfectly consistent with Landau's theory as discussed for general thermodynamical systems.  Because of this analogy we expect that all other critical exponents are the same as those calculated in the Landau's approach \cite{8}.

\section{Summary and conclusions}
In this paper we discussed some nuclear equations of state bridging basic properties of nuclear matter near its ground state and a possible second order phase transition to the quark-gluon plasma at high densities.  We have determined a critical density of about three times normal ground state density for the QGP in the case where
a critical value $\delta=3$, this corresponds to the so called mean field value or classical value of the critical exponent\cite{8}. To go beyond this classical value we used Landau's theory of phase transition and for $\delta=5$ we determined an EOS to $O(\rho^6)$ which gives a critical density about five times the ground state density of nuclear matter.
Using an MIT bag model, it is possible to estimate a critical density for the QGP of about five times normal nuclear matter density using a bag constant $B^{1/4}=206 $MeV\cite{16}. Other estimates using phenomenological hadronic and QGP EOS give similar densities but a first order phase transition\cite{Csernai}.
In order to see if there is a first or second order phase transition or a simple cross-over from one state to the other, reliable experimental data in the region at high baryonic densities and relatively small temperatures are needed or data from heavy stars compared to refined theoretical models.  Such quantities could be for instance collective flow compared to microscopic
calculations which implement the EOS discussed here.  As we have seen the pressure at high densities is completely different in the EOS with or without a QGP phase transition.  In particular if the transition is second order than the corresponding EOS cannot be much different from the one estimated here when $\delta=5$. 
At lower densities a liquid gas phase transition might occur at the critical density $\rho_c=0.37\rho_0$  and  $T_c=18$ MeV. Such values are rather independent from the EOS chosen when fixing the compressibility.  However the critical temperature depends somewhat on the assumed classical or quantum statistics.  In the region of densities and
temperature of the LG transition it seems that there is no reason why a classical approximation should work.  However, presently most experimental and theoretical results in this region are obtained using classical methods.  This is a feature that should be improved in future works in order to have more reliable constraints
on the EOS\cite{15a}.

Finally, we found that for odd orders as, $O(\rho^3)$, $O(\rho^5)$ this approach is not suitable to describe the basic properties of our system.  The even order
 $O(\rho^4)$, $O(\rho^6)$ approximation result in  more acceptable behavior of the energy per particle as function of density. It could be possible that the collective character in regions with high density of particles are associated with even numbers \cite{18,19}.

\begin{acknowledgments}One of us (RM) would like to acknowledge the support by DOE, NSF-REU Program, and the support of many people from  Texas A\&M Cyclotron Institute and Instituto de Ciencias Nucleares(ICN)-UNAM. We thank prof. J.Natowitz for discussions.
\end{acknowledgments}


\begin{thebibliography}{99}
\bibitem{1}
A. Bonasera, F. Gulminelli and J. Molitoris,  
Phys. Rep.{\bf 243}, 1 (1994).
\bibitem{2}
G. Bertsch and S. Dasgupta
Phys. Rep.{\bf 160}, 189 (1988).
\bibitem{Csernai}
L.P. Csernai Introduction to Relativistic Heavy Ion Collisions ( Wiley, New York 1994).
\bibitem{3}
W. D. Myeres,  \textit{Atomic Data Nucl. Data Tables}, \textbf{17}, 411 (1978).
\bibitem{4}
H. A. Bethe, \textit{Ann. Rev. Nucl. Sci.} \textbf{21} 93 (1971).
\bibitem{5}
J. P. Blaizot, D. Gogny and B Grammaticos, Nucl. Phys. \textbf{A265}, 315 (1976).
\bibitem{shlomo}
B.K.Agrawal, S.Shlomo and V.Kim Au, Phys Rev. \textbf{C72}, 014310 (2005).
\bibitem{8}
Huang K., Statistical Mechanics (J. Wiley and Sons, New York) 1987,  2n ed.;\\
Landau L and Lifshits F. Statistical Physics (Pergarmon, New York) 1980.
\bibitem{albergo} S.Albergo et al., Nuovo Cimento \textbf{89}, 1(1985).
\bibitem{pocho}J.Pochodzalla et al. PRL\textbf{75}, 1040(1995).
\bibitem{15}J.B.Natowitz et al., Phys. Rev.\textbf{C65}, 034618(2002);\\
S. Wuenschel  et al., Nucl. Phys.\textbf{A843}, 1 (2010).
\bibitem{15a}J.B.Natowitz et al., Phys. Rev.Lett.\textbf{104}, 202501(2010).
\bibitem{10}
A. Bonasera et al., Rivista del Nuovo Cimento \textbf{23},  1(2000).
\bibitem{11}
T. S. Olson and W. A. Hiscock, Phys Rev.\textbf{C39}, 1818 (1989).
\bibitem{12}
C. Grant and J. Kapusta, Phys. Rev. \textbf{C32}, 663 (1985);\\
 A. Goodman, J.Kapusta and A.Z.Mekijan, Phys. Rev.\textbf{C30}, 851 (1984).
\bibitem{13}
A. Bonasera and M. Di Toro, Lettere at Nuovo Cimento \textbf{44}, 172 (1985).
\bibitem{16}
C.Y. Wong, Introduction to High-Energy Heavy Ion Collisions (World Scientific Singapore 1994).
\bibitem{17}
A. Bonasera, Phys. Rev.\textbf{C62},  O52202(R)(2000);\\
A.Bonasera, Nucl.Phys.\textbf{A681}, 64c(2001);\\
S.Terranova and A.Bonasera, Phys. Rev.\textbf{C70},  O24906(2004);\\
S.Terranova, D.M.Zhou and A.Bonasera, Eur.Phys.J.\textbf{A26},  333(2005);\\
Z.G.Tan and A.Bonasera, Nucl.Phys.\textbf{A784}, 368(2007).
\bibitem{Landau-Vol6}
Landau L and Lifshits F. Fluid Mechanics (Pergarmon, New York) 1980. 
\bibitem{18}
A. Arima and F. Iachello, Phys. Rev. Lett.\textbf{35}, 1069 (1975).
\bibitem{19}
A. Bohr and B.R. Mottelson, Mat. Phys. Dan. Vid. Selks. \textbf{27} No 16 (1953).
\newpage
\end{thebibliography}
\end{document}